\documentclass[aps,prl,unsortedaddress,twocolumn,showpacs]{revtex4-1}

\usepackage{amsfonts}

\usepackage[pdftex]{graphicx}
\usepackage{epstopdf}
\usepackage{amsmath}
\usepackage{amssymb}
\usepackage{verbatim}
\usepackage[ulem=normalem]{changes}
\usepackage{booktabs}
\usepackage{xcolor}

\usepackage{soul}

\usepackage[T1]{fontenc}

\newcommand{\ra}[1]{\renewcommand{\arraystretch}{#1}}

\begin{document}

\title{Adiabatic processes realized with a trapped Brownian particle}

\author{Ignacio A. Mart\'inez$^{1,2}$, \'Edgar Rold\'an$^{1,3,4}$, Luis Dinis$^{4,5}$, Dmitri Petrov$^{1,6}$\medskip}
\author{Ra\'ul A. Rica$^1$}\email{Corresponding author: rul@ugr.es}
\affiliation{$^1$ICFO $-$ Institut de Ci\`encies Fot\`oniques, Mediterranean Technology Park, 08860, Castelldefels (Barcelona), Spain. \\
$^2$ Laboratoire de Physique, \'Ecole Normale Sup\'erieure, CNRS UMR5672 46 All\'ee d'Italie, 69364 Lyon, France.\\
$^3$ Max Planck Institute for the Physics of Complex Systems, N\"othnitzerstrasse 38, 01187 Dresden, Germany. \\
$^4$GISC $-$ Grupo Interdisciplinar de Sistemas Complejos. Madrid, Spain.\\
$^5$Departamento de F\'isica At\'omica, Molecular y Nuclear, Universidad Complutense de Madrid, 28040,  Madrid, Spain. \\
$^6$ICREA $-$ Instituci\'o Catalana de Recerca i Estudis Avan\c cats, 08010, Barcelona, Spain}

\begin{abstract}

We experimentally realize quasistatic adiabatic processes using a single optically-trapped microsphere immersed in water whose effective temperature is controlled by an external random electric field.  A full energetic characterization of adiabatic processes that preserve either the position distribution or the full phase space volume is presented. We show that only in the latter case the exchanged heat and the change in the entropy of the particle vanish when averaging over many repetitions.  We provide analytical expressions for the distributions of the fluctuating heat and entropy, which we verify experimentally. We show that the heat distribution is asymmetric for any non-isothermal quasistatic process.  Moreover, the  shape of the distribution of the system entropy change in the adiabatic processes depends significantly on the number of degrees of freedom that are considered for the calculation of system entropy. 

\end{abstract}


\maketitle



Stochastic energetics~\cite{SekimotoPTPS1998,SekimotoBook2010} and the fluctuation theorems\textbf~\cite{seifert2012stochastic} have been developed as the theoretical framework that studies thermodynamics at small scales, thus establishing the emerging field of stochastic thermodynamics. In parallel, recent advances on micromanipulation and force-sensing techniques~\cite{ciliberto2010fluctuations} have allowed to measure the dynamics and energy changes in physical systems where thermal fluctuations are relevant~\cite{DuckerNature1991,VisscherNature1999,BustamantePhysToday2005,gieseler2014dynamic,millen2013nanoscale} and to test theoretical results derived from stochastic thermodynamics~\cite{CilibertoJphys1998,LiphardtScience2002,wang2002experimental,CollinNature2005,WangPRE2005,ToyabeNatPhys2010}. As a major application, miniaturization of thermodynamic engines to single-molecule devices has been possible for the case of Stirling engine~\cite{Blickle2011} or a variety of Maxwell's demons~\cite{ToyabeNatPhys2010,Roldan2014Universal,koski2014experimental}.

Until now, the design of microscopic heat engines has been restricted to those cycles formed by isothermal processes or instantaneous temperature changes~\cite{Blickle2011}, where the validity of a heat fluctuation theorem has been tested~\cite{gomez2011heat}. Recent works have shown that exerting random forces on a microscopic particle one can accurately tune the effective kinetic temperature of the particle both under equilibrium~\cite{gomez2010steady,Martinez2013,berut2014energy} and nonequilibrium driving~\cite{mestres2014realization}. However, the application of such a technique to implement non-isothermal processes has not been fully exploited yet~\cite{roldan2014measuring}. 

Among all the non-isothermal processes, adiabatic processes are of major importance in thermodynamics since they are the building blocks of the Carnot engine~\cite{carnot1986reflexions}. \emph{Microadiabaticity}, i.e. adiabaticity at the microscopic scale, cannot be realized for single-trajectories due to the unavoidable heat flows between microscopic systems and their surroundings. However,  a process where no net heat transfer is obtained when averaged over many trajectories could in principle be realized. Although several theoretical proposals are available~\cite{sekimoto2000carnot,Schmiedl2008,lahiri2012fluctuation,Bo2013}, their experimental implementation is still lacking.

\begin{figure}
\includegraphics[width=6cm]{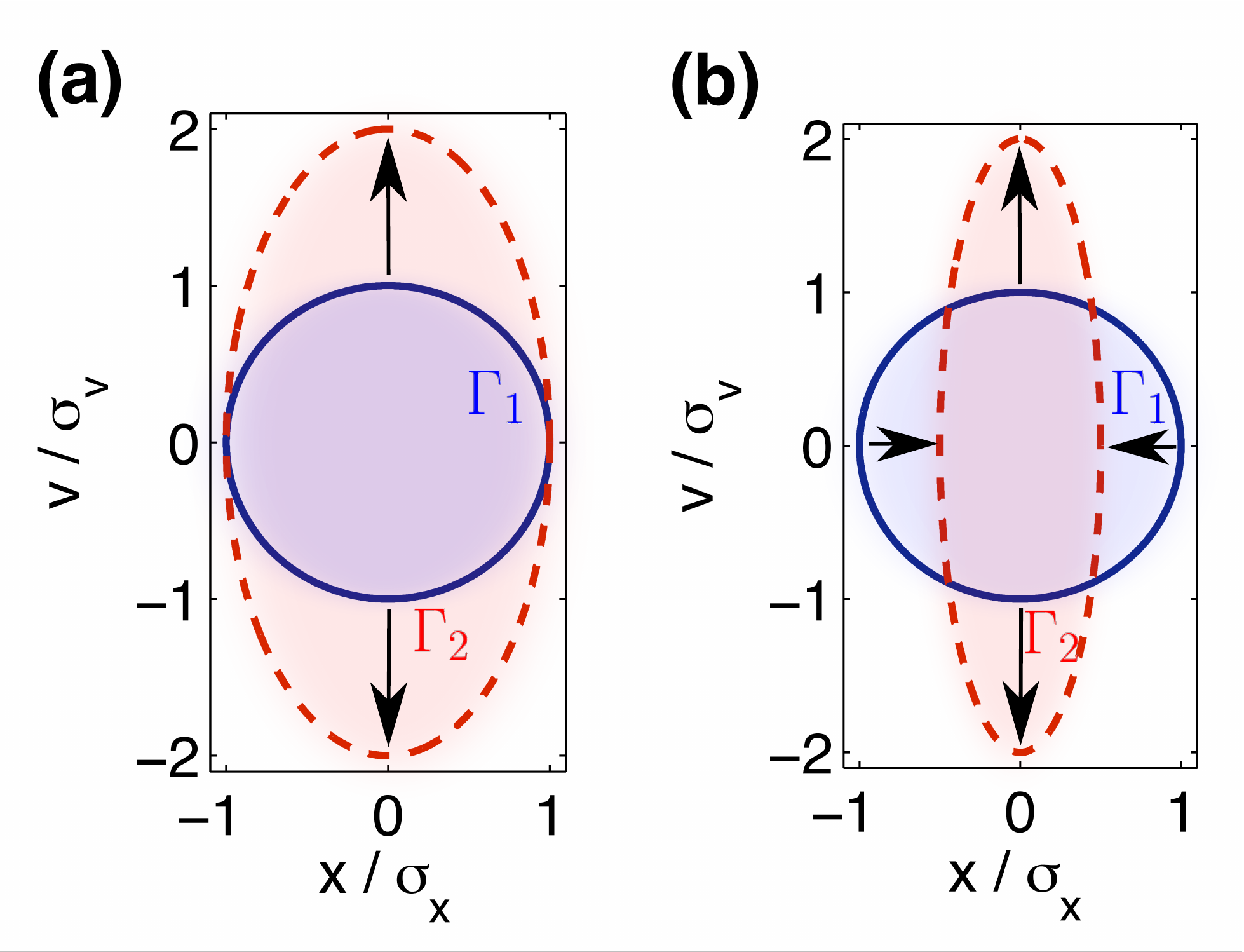} 
\caption{Illustration of the pseudo-adiabatic and adiabatic processes. A Brownian particle of mass $m$ is immersed in a thermal bath at temperature $T$, moves in one dimension $x$ with velocity $v$ and is trapped with a harmonic potential $U(x)=\frac{1}{2}\kappa x^2$. The blue solid circle $\Gamma_1$ represents the ensemble of microstates described by a Hamiltonian  $\mathcal{H}(x,v)= \frac{1}{2}\kappa x^2 + \frac{1}{2}mv^2$ with a given energy $\mathcal{H}(x,v)=E=kT$. The units of position and velocity are normalized by their standard deviation from equipartition theorem, $\sigma_x=\sqrt{kT/\kappa}$ and $\sigma_v=\sqrt{kT/m}$. The red dashed ellipse $\Gamma_2$ is the microstate set at the same energy but after two different adiabatic processes: (a) Pseudo-adiabatic process, where $T_{\rm fin}=2T$ and $\kappa_{\rm fin}=2\kappa$ (i.e., $T/\kappa=\rm{const}$);  (b) Adiabatic process, where $T_{\rm fin} = 2T$ and $\kappa_{\rm fin} = 4\kappa$ (i.e., $T^2/\kappa=\rm{const}$). The arrows indicate the direction in which the process occurs. Notice that the area of phase space that satisfies $\mathcal{H}(x,v)\leq E$ is conserved only along the adiabatic process.
\label{fig:huevoscoloridos}}
\end{figure}

In this Letter, we report on the realization of quasistatic adiabatic processes with an optically-trapped microparticle immersed in water whose kinetic temperature is controlled by means of an external noisy electric field~\cite{Martinez2013,roldan2014measuring}. We provide a complete characterization of the thermodynamics of such adiabatic processes. The contributions due to the heat transferred to the momentum degree of freedom are also considered~\cite{roldan2014measuring}, thus adopting the full, underdamped description of the system.  Interestingly, we show that doing so does not constitute a trivial extension of the overdamped description, but distinct features clearly arise. In particular, we discuss the shape of the distribution of the fluctuations of heat and entropy in both descriptions, concluding that asymmetries in the heat distributions are a fingerprint of quasistatic non-isothermal processes. 

In classical Hamiltonian systems, the total heat $Q$ transferred in quasistatic adiabatic processes vanishes, and the heat distribution is $\rho_{\mathcal{H}}(Q) = \delta (Q)$. The work, $W=\Delta U$, $U$ being the internal energy, is exponentially distributed. In the microscopic regime, one can attain processes where $\langle Q \rangle = 0$, $\langle \cdot\rangle$ denoting average over many realizations in the quasistatic limit. In the latter case, at odds with the Hamiltonian case, the work is delta distributed, $\rho(W)= \delta (\Delta U + \langle Q\rangle)$~\cite{SekimotoBook2010} whereas the heat is exponentially distributed with the same distribution as $\rho_{\mathcal{H}}(W)$, as shown in the Supplemental Material~\cite{supplementary}.  Microadiabatic processes are those where the phase space volume is conserved~\cite{sekimoto2000carnot}. In the overdamped limit, where changes in the momentum degree of freedom are neglected, such condition is met by keeping the position distribution constant~\cite{seifert2005entropy}. However, as we discuss below, the overdamped approximation is incomplete when dealing with non-isothermal processes, and a full underdamped description is mandatory. A process where the position distribution is conserved is therefore a {\em pseudo-adiabatic}, since an unavoidable amount of heat is transferred due to the kinetic energy change~\cite{Schmiedl2008}. In contrast, in the actual \emph{adiabatic} process, the full phase space (position and momentum) volume is conserved and no net heat is transferred to the particle~\cite{sekimoto2000carnot,Bo2013}. Figure~\ref{fig:huevoscoloridos} illustrates the difference between the evolution of the phase space along both quasistatic pseudo-adiabatic and adiabatic processes for a Brownian particle trapped with a harmonic potential. Notice that only in the adiabatic process the phase space volume enclosed by the energy surface defined by the system's energy at every moment $\mathcal{H}(x,v)=E$ is conserved, as required for a quasistatic and adiabatic change of parameters~\cite{sekimoto2000carnot}.

Our system of study is a microparticle of radius $R=500 \rm nm$ immersed in water trapped by an optical harmonic optical potential $U(x)=\frac{1}{2}\kappa x^2$, where $\kappa$ is the stiffness of the trap and $x$ the position of the particle with respect to the trap center. The key capability of our setup is the independent control of the kinetic temperature of the trapped bead $T_{\rm kin}$ and $\kappa$, thus allowing one to design a large variety of different thermodynamic processes~\cite{roldan2014measuring,mestres2014realization}. Both parameters can be electronically synchronized in order to fullfill any desired protocole with high time resolution,  of the order of $\mu\rm s$.

\begin{figure}
\includegraphics[width=8.5cm]{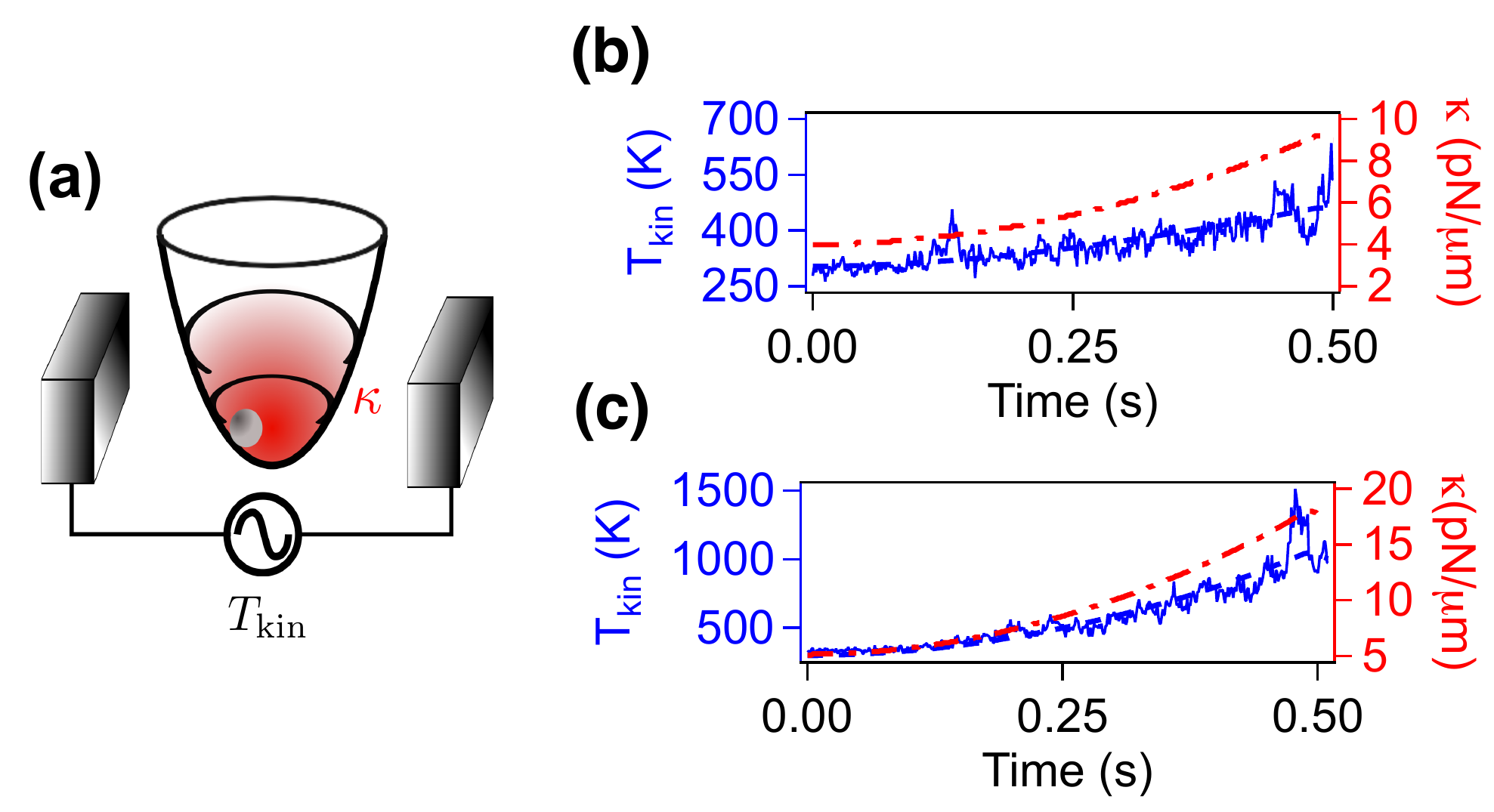} 
\caption{Experimental setup and experimental protocols. (a) Sketch of the experimental setup. The kinetic temperature $T_{\rm{kin}}$ of a micro particle in an optical trap of stiffness $\kappa$ is controlled with a noisy electric field. (b) Pseudo-adiabatic protocol. Kinetic temperature from the mean squared displacement, $\kappa \langle x^2\rangle /k$ (left axis, blue solid line), kinetic temperature from the calibration (left axis, blue dashed line) and stiffness of the trap (right axis, red dash-dot line) as functions of time. (c) The same for the adiabatic process. Notice the larger fluctuations as $T_{\rm {kin}}$ increases.}
\label{fig:protocols}
\end{figure}

$T_{\rm kin}$ is defined from the application of equipartition theorem to the fluctuations of the position of the bead in the trap as follows. These fluctuations obey equipartition theorem, $\kappa \langle x^2 \rangle = kT$, $k$ being Boltzmann's constant and $T$ the temperature of the sample~\cite{greiner1999thermodynamics}. Applying to the particle an external random force characterized by a Gaussian white noise process of amplitude $\sigma$, we can mimic the kicks of the solvent molecules to the bead in a higher temperature reservoir $T_{\rm kin}$. 
From equipartition theorem, the kinetic temperature of the particle depends on its mean squared displacement 
$\langle x^2\rangle$, $T_{\rm kin}=\kappa \langle x^2 \rangle/k=T+\sigma^2/2\gamma k\ge T$, $\gamma=6\pi\eta R$ being the Stokes friction on a sphere of radius $R$ in a fluid with kinetic viscosity $\eta$ far away from a surface. These two parameters can be easily controlled, since $\kappa$ is proportional to the intensity of the trapping laser \cite{mazolli2003theory} and $T_{\rm kin}$ increases linearly with the square of the amplitude of a noisy voltage applied to a pair of electrodes in the fluid chamber~\cite{supplementary,martinez2014noise}. See Fig.~\ref{fig:protocols}(a) for a sketch of the experimental system.

Following the usual approach, we first implement a pseudo-adiabatic protocol where the entropy is conserved in the overdamped approximation, i.e. where $T_{\rm kin}(t) / \kappa(t) = \text{const}$ (see Supplemental Material~\cite{supplementary} for a proof), as the one shown in Fig.~\ref{fig:protocols}(b). An actual microadiabatic process is achieved making $T_{\rm kin}^2(t) / \kappa(t) = \text{const}$ (see Supplemental Material~\cite{supplementary})  and is implemented as shown in Fig.~\ref{fig:protocols}(c). All the protocols presented here have a duration of $\tau=0.5\,\ \rm s$. Since the relaxation time of the particle in the trap $\tau_c$ is of the order of milliseconds~\cite{roldan2014measuring}, then $\tau \gg \tau_c \sim \rm ms$ and the processes can be considered as quasistatic. The latter is confirmed in Figs.~\ref{fig:protocols}(b-c), where we show that the measured kinetic temperature fluctuates around the value prescribed by the protocol. 

After defining and implementing the desired protocols, we calculate the thermodynamic quantities from measurements of the position of the trapped bead the stiffness of the trap and the kinetic temperature of the bead, the latter being obtained from standard calibration procedures~\cite{supplementary,martinez2014noise}. The data acquisition frequency was $f=1/2\pi\Delta t=1\,\rm kHz$.  The work done on the particle in the time interval $[t,t+\Delta t]$ is calculated as $\delta W (t) = U(x_t,t+\Delta t) - U(x_t,t)$, $x_t$ being the position of the particle at time $t$~\cite{Roldan2014Universal}. The heat transferred from the thermal bath to the position of the particle is calculated as $\delta Q_x(t) = U(x_{t+\Delta t},t+\Delta t) - U(x_t,t+\Delta t)$. The internal energy change is measured as the sum of the heat and the work transferred to the particle, $\Delta U (t) = \delta W (t)+\delta Q_x(t)= U(x_{t+\Delta t},t+\Delta t) - U(x_{t},t)$. In the limit $\Delta t\to 0$, the cumulative sum up to time $t$ of our definitions of heat and work return Sekimoto's expressions $\int \delta W(t) \to \int \frac{\partial U}{\partial t} dt$ and  $\int \delta Q_x(t) \to \int \frac{\partial U}{\partial x}\circ dx$~\cite{SekimotoBook2010}. Ensemble averages and probability distributions are calculated from datasets of $900$ repetitions of each process.

We estimate the kinetic energy changes following the technique described in~\cite{roldan2014measuring}. The sampling frequency in our experiment is far below the momentum relaxation frequency $f_p = \gamma/m \sim \text{MHz}$, $m$ being the mass of the bead ~\cite{kheifets2014observation}. Therefore, we can only measure time averaged velocities $\overline v_t = (x_{t+\Delta t} - x_t)/\Delta t$ rather than instantaneous velocities  $v_t$. In the quasistatic limit, we can obtain the mean squared instantaneous velocity from the mean squared time averaged velocity, $\langle v_t^2\rangle = \mathcal{L}_t \langle \overline{v}_t^2\rangle$, where $\mathcal{L}_t = \mathcal{L}_t (f,\kappa_t,\gamma,m)$ is a function of the sampling frequency as well as of the parameters of the system at time $t$ (stiffness, mass, friction coefficient)~\cite{roldan2014measuring}. The ensemble average kinetic energy change can be therefore calculated as $\langle \Delta E_{\rm kin} (t)\rangle  = \frac{m}{2} [\langle v_{t+\Delta t}^2\rangle -\langle v_t^2\rangle]= \frac{m}{2} [ \mathcal{L}_{t+\Delta t} \langle \overline{v}_{t+\Delta t}^2\rangle - \mathcal{L}_t \langle \overline{v}_t^2\rangle]$. In addition, we can assess the distribution of the instantaneous velocity from the distribution of the time averaged velocity. The latter is Gaussian with zero mean and the variance is related to that of the velocity distribution by $\sigma^2 (v_t) = \mathcal{L}_t \sigma^2 (\overline v_t)$, allowing to reconstruct the velocity distribution as~$\rho(v_t) =  \rho(\sqrt{\mathcal{L}_t}\overline v_t)$, as shown in the Supplemental Material~\cite{supplementary}.

The velocity distributions can be exploited to determine the entropy of the system in the underdamped description at any time $t$, $S_t = -k\ln \rho(x_t,v_t,t)$, or equivalently, the average system entropy change in the interval $[t,t+\Delta t]$, $\langle \Delta S(t)\rangle = \langle S_{t+\Delta t} \rangle - \langle S_{t} \rangle$, where $\langle S_t\rangle  = -k\int \rho(x_t,v_t,t) \ln \rho(x_t,v_t,t)\, dx_t\, dv_t$ is the ensemble average system entropy at time $t$~\cite{seifert2005entropy,hilbert2014thermodynamic}. We also consider the overdamped system entropy change obtained when we neglect the velocity degree of freedom, $\langle \Delta S_x(t)\rangle = \langle S_{x,t+\Delta t} \rangle - \langle S_{x,t} \rangle$, $\langle S_{x,t}\rangle  = -k\int \rho(x_t,t) \ln \rho(x_t,t)\, dx_t$ being the entropy of the system in the position degree of freedom.

\begin{figure}
\includegraphics[width=8.5cm]{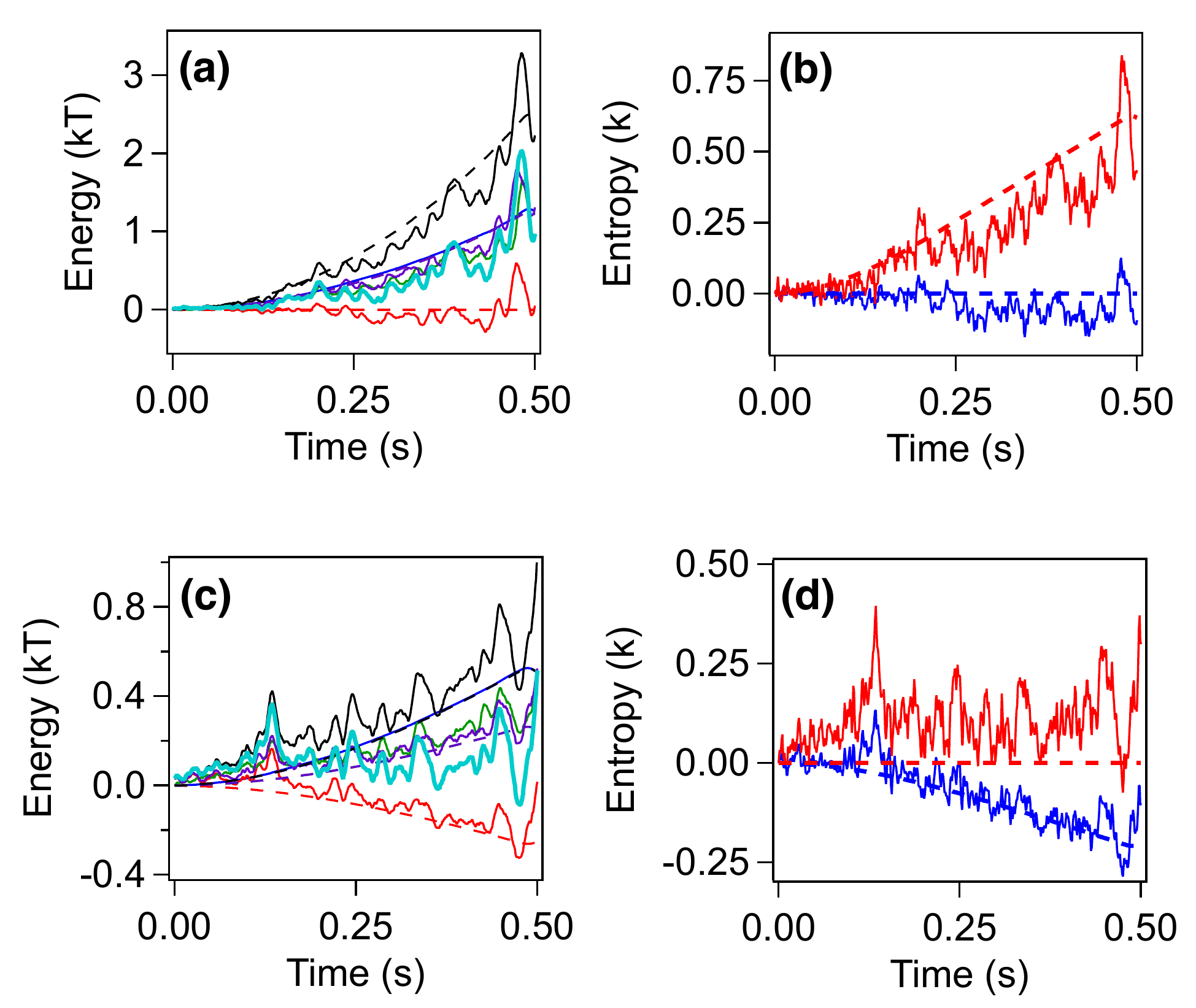} 
\caption{Ensemble averages of the cumulative sums of thermodynamic quantities as a function of time in the pseudo-adiabatic (a-b) and adiabatic (c-d) processes. (a) Energy as a function of time for the pseudo-adiabatic process, $\langle W(t) \rangle$ (blue),  $\langle Q_x(t) \rangle$ (red), $\langle \Delta E_{\rm kin}(t) \rangle$ (green), $\langle Q(t) \rangle$ (cyan), $\langle  \Delta U(t) \rangle$ (magenta) and $\langle  \Delta E(t) \rangle$ (magenta). (b) System entropy as a function of time for the pseudo-adiabatic process, $\langle \Delta S_{x}(t) \rangle$ (blue) and $\langle \Delta S(t) \rangle$ (red).  (c) Energetics of the adiabatic process. (d) System entropy change in the adiabatic process. $\langle \Delta S_{x}(t) \rangle$ (blue) and $\langle \Delta S(t) \rangle$ (red). Ensemble averages are obtained from $900$ repetitions of cycles of duration $\tau=0.5\,\rm s$ using a sampling rate of $f=1\,\rm kHz$. Dashed curves are the theoretical values of the thermodynamic quantities obtained in the quasistatic limit. }
\label{fig:termoadiabatics}
\end{figure}

Using the aforementioned definitions of energetic quantities and entropy, we can now characterize the two types of microadiabatic processes. Let us first analyse the pseudo-adiabatic process. Figure~\ref{fig:termoadiabatics}(a) shows ensemble averages of the cumulative sum of work, heat, kinetic energy, internal energy and total energy, which coincide with the expected values from equilibrium thermodynamics. The average heat transferred to the position degree of freedom vanishes within experimental errors, yielding a net positive total value of the heat $\langle Q (t)\rangle = \langle Q_x (t) \rangle +  \langle \Delta E_{\rm kin} (t) \rangle =  \frac{k}{2}[T_{\rm kin,t} - T_{\rm kin,0}]>0$. The average overdamped entropy change vanishes along the protocol, $\langle \Delta S_{x} (t)\rangle=0$, as shown in Fig.~\ref{fig:termoadiabatics} (b).  The pseudo-adiabatic nature of the protocol is revealed as a positive full system entropy change, $\langle\Delta S (t)\rangle >0$ [red curve in Fig.~\ref{fig:termoadiabatics} (b)]. For the adiabatic protocol [Fig.~\ref{fig:protocols}(c)], the ensemble average of the {\em total} heat transferred to the particle vanishes within experimental errors, $\langle Q (t)\rangle =0$, as shown in Fig.~\ref{fig:termoadiabatics} (c). As a result, the system entropy change vanishes along the adiabatic process $\langle\Delta S(t)\rangle = 0$ despite entropy is reduced in the position degree of freedom, as shown in Fig.~\ref{fig:termoadiabatics} (d).

\begin{figure}
\includegraphics[width=5.5cm]{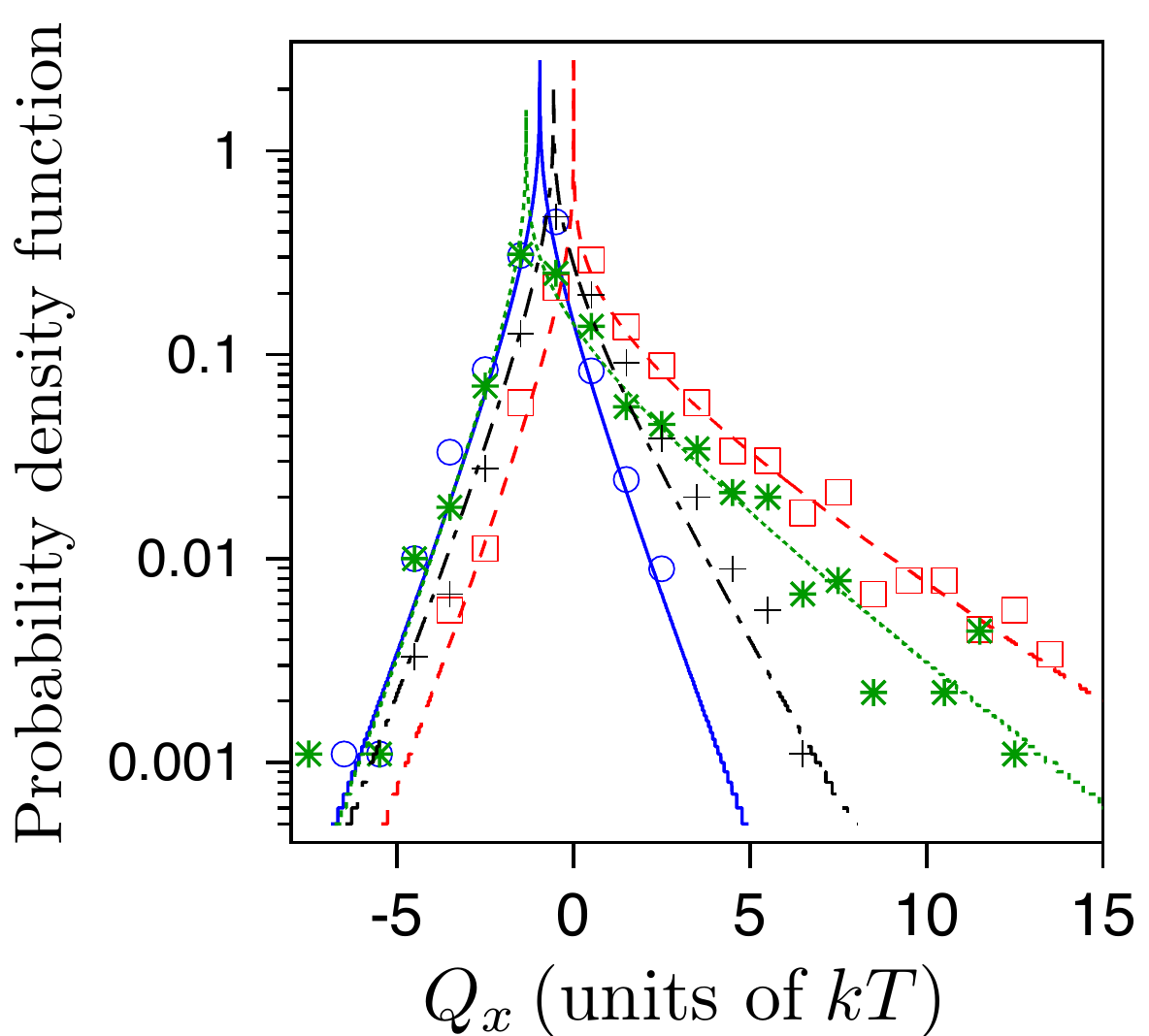} 
\caption{Distribution of the heat absorbed by the particle in the position degree of freedom $Q_x$ for different thermodynamic processes: isothermal (blue circles), isochoric (red squares), pseudo-adiabatic (green stars) and adiabatic (black crosses). The distributions are obtained from $900$ cycles of duration $\tau = 0.5\,\rm s$ each. The lines are theoretical distributions obtained from Eq.~\eqref{eq:PDFQ} using the initial and final values of kinetic temperature and stiffness used in the experiments, for the different processes:  isothermal (solid line), isochoric (dashed line), pseudo-adiabatic (dotted line) and adiabatic (dashed-dotted line).}
\label{fig:heatdistributions}
\end{figure}

Let us now consider the fluctuations of the measured quantities. Again, their values fluctuate around the predictions obtained in the quasistatic limit. The shape of the these distributions reveals qualitative differences between the considered processes. First, in Fig.~\ref{fig:heatdistributions} we show the experimental probability density function of $Q_x$ (symbols) for the two adiabatic processes considered. We also include the distributions for two control processes: an isothermal process where $T_{\rm kin, t}= 300\,\rm K$ and $\kappa$ changes linearly in time from $\kappa_0=(5.0\pm 0.2)\text{pN}/\mu\text{m}$ to $\kappa_\tau = (28.0\pm 0.2)\text{pN}/\mu\text{m}$, and an isochoric processes where $\kappa_t=(18.0\pm 0.2)\text{pN}/\mu\text{m}$ and the kinetic temperature changes linearly from $T_{\rm kin, 0}=300\,\rm K$ to $T_{\rm kin, \tau}=1200\,\rm K$~\cite{roldan2014measuring}.

Remarkably, we notice that the heat distribution is asymmetric around its mean for all the non-isothermal processes. The measured heat distributions can be well described by
\begin{equation}
\rho(Q_x)=\frac{\beta_G}{\pi} \exp\left[-\frac{\Delta \beta}{2} (Q_x+\langle W\rangle)\right] K_0 \left[\overline{\beta}|Q_x+\langle W\rangle|\right],
\label{eq:PDFQ}
\end{equation}
where $\beta_0=1/kT_{\rm kin} (0)$, $\beta_{\tau} = 1/kT_{\rm kin} (\tau)$, $\Delta \beta = \beta_\tau - \beta_0$, $\overline{\beta} = \frac{\beta_0 + \beta_\tau}{2}$, $\beta_G=\sqrt{\beta_0\beta_\tau}$, $K_0$ is zeroth order modified Bessel function of the second kind and $\langle W\rangle$ is the ensemble average of the work in the quasistatic limit~\cite{supplementary}. Equation~\eqref{eq:PDFQ} was obtained with the only assumption of quasistaticity along the process, and proves that the asymmetry of the distribution of $Q_x$ around $-\langle W\rangle$ is a consequence of the non-isothermal character of the process, and not of any nonequilibrium constraint of the system, as suggested in Ref.~\cite{gomez2011heat}. For the isothermal case, $\Delta \beta=0$ and we recover the symmetric distribution $\rho(Q_x)=\frac{\beta}{\pi} K_0 \left[\beta|Q_x+\langle W\rangle|\right]$ firstly derived by Imparato {\em et al}~\cite{imparato2007work}.

\begin{figure}
\includegraphics[width=5.5cm]{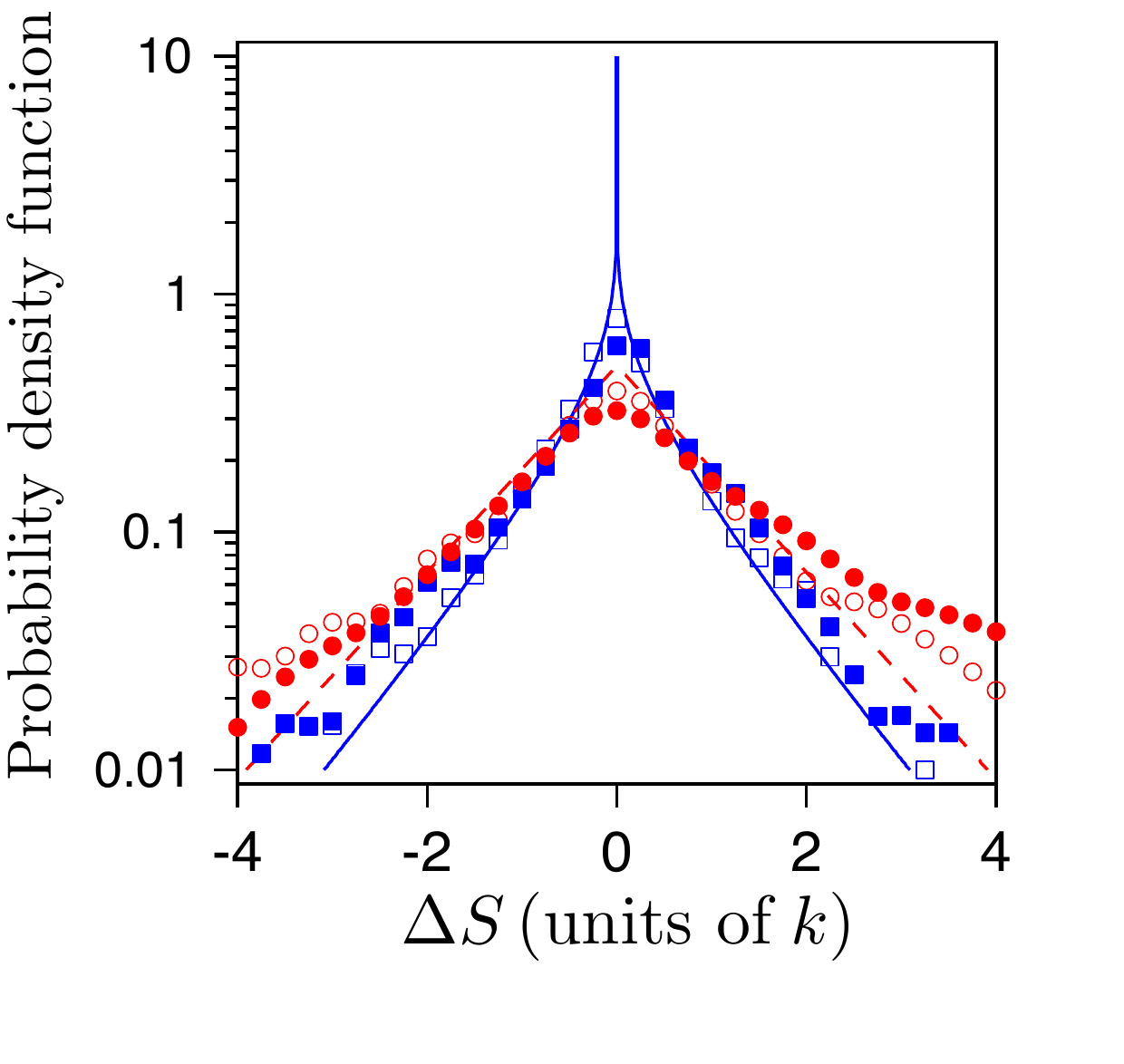} 
\caption{Distribution of the system entropy change in the overdamped description, $\Delta S_{x}$ (blue squares) and of the total system entropy change $\Delta S$ (red circles) in the pseudo-adiabatic (open symbols) and adiabatic (closed symbols) processes shown in Fig.~\ref{fig:protocols} (b) and (c). The distributions are obtained from $900$ cycles of duration $\tau = 0.5\,\rm s$ each. Theoretical distributions for $\Delta S_x$ [blue solid curve, Eq. \eqref{eq:rhoSx}] and  $\Delta S$ [red dashed curve, Eq. \eqref{eq:rhoS}]  are also shown. All the quantities are shifted by their mean such that the mean of the represented quantities is zero.}
\label{fig:distr_entropy}
\end{figure}

The asymmetry observed in the heat fluctuations is not present in the distribution of the entropy. The distribution of the overdamped entropy change along the whole process, $\Delta S_x (\tau) = S_x (\tau) - S_x(0)$ is symmetric around around its mean value for both pseudo-adiabatic and adiabatic cases, as shown in Fig.~\ref{fig:distr_entropy}. Both distributions fit well to the expected value for general quasistatic non-isothermal processes~\cite{supplementary},
\begin{equation}
\rho(\Delta S_x) =\frac{1}{\pi k} K_0 \left(\frac{|\Delta S_x-\langle \Delta S_x\rangle|}{k}\right).
\label{eq:rhoSx}
\end{equation}
We also calculate the distribution of the full system entropy change along the whole process, $\Delta S (\tau) = S(\tau) - S(0)$ in both pseudo-adiabatic and adiabatic processes (see Fig.~\ref{fig:distr_entropy}). System entropy change is distributed symmetrically around its mean value but presents a different qualitative behavior, in this case described by~\cite{supplementary}:
\begin{equation}
\rho(\Delta S) = \frac{1}{2k}\exp\left(-\frac{|\Delta S-\langle\Delta S\rangle|}{k}\right)
\label{eq:rhoS}
\end{equation}
Notice that in the case of the full system entropy change, the agreement with the theory extends over one order of magnitude less than in the overdamped description, $\Delta S_x$. This mismatch is caused by the poor estimation of the tails of the distribution of the instantaneous velocity from the distribution of the time averaged velocity.

To summarize, we have realized quasistatic adiabatic processes with a single microparticle trapped with optical tweezers. We have studied the difference between the pseudo-adiabatic (position distribution conserving) and adiabatic (phase space volume conserving) processes, showing that only the latter are such that the average total heat vanishes in the ensemble average. The fluctuations of the heat transferred to the position of the particle have been shown to be asymmetric for any non-isothermal (equilibrium or nonequilibrium) thermodynamic process. The description of the dynamics of the system with full or limited information affects not only to the average values of the entropy but also to the fluctuations, showing a different qualitative behavior. The microadiabatic protocols studied in the present work could be used to design a microscopic-sized Carnot engine by a cyclic sequence of isothermal and adiabatic processes, thus extending our understanding of micro and nano electromechanical systems towards new and efficient engines~\cite{sekimoto2000carnot,Schmiedl2008,sanchez2010optimization,Bo2013,rana2014single}.

\medskip

We acknowledge enlightening theoretical discussions with J. M. R. Parrondo.
I.A.M., E.R., D.P. and R.A.R. acknowledge financial support from the Fundaci\'o Privada Cellex
Barcelona, Generalitat de Catalunya grant 2009-SGR-159, and from the Spanish
Ministry of Science and Innovation (MICINN FIS2011-24409).
E.R. and L.D.  acknowledge financial support from ENFASIS (Spanish Government).
The initial ideas of this work were conceived by Prof. D. Petrov, leader of the Optical Tweezers group at ICFO, who passed away on 3rd February 2014.


%


\newpage\newpage

\renewcommand{\thefigure}{S\arabic{figure}}
\renewcommand{\theequation}{S\arabic{equation}}
\renewcommand{\thetable}{S\arabic{table}}
\setcounter{figure}{0}
\setcounter{equation}{0}
\setcounter{table}{0}

\newpage \newpage

\section{Experimental setup}
\label{sec:experimentalsetup}

\begin{figure}
\includegraphics[width=9cm]{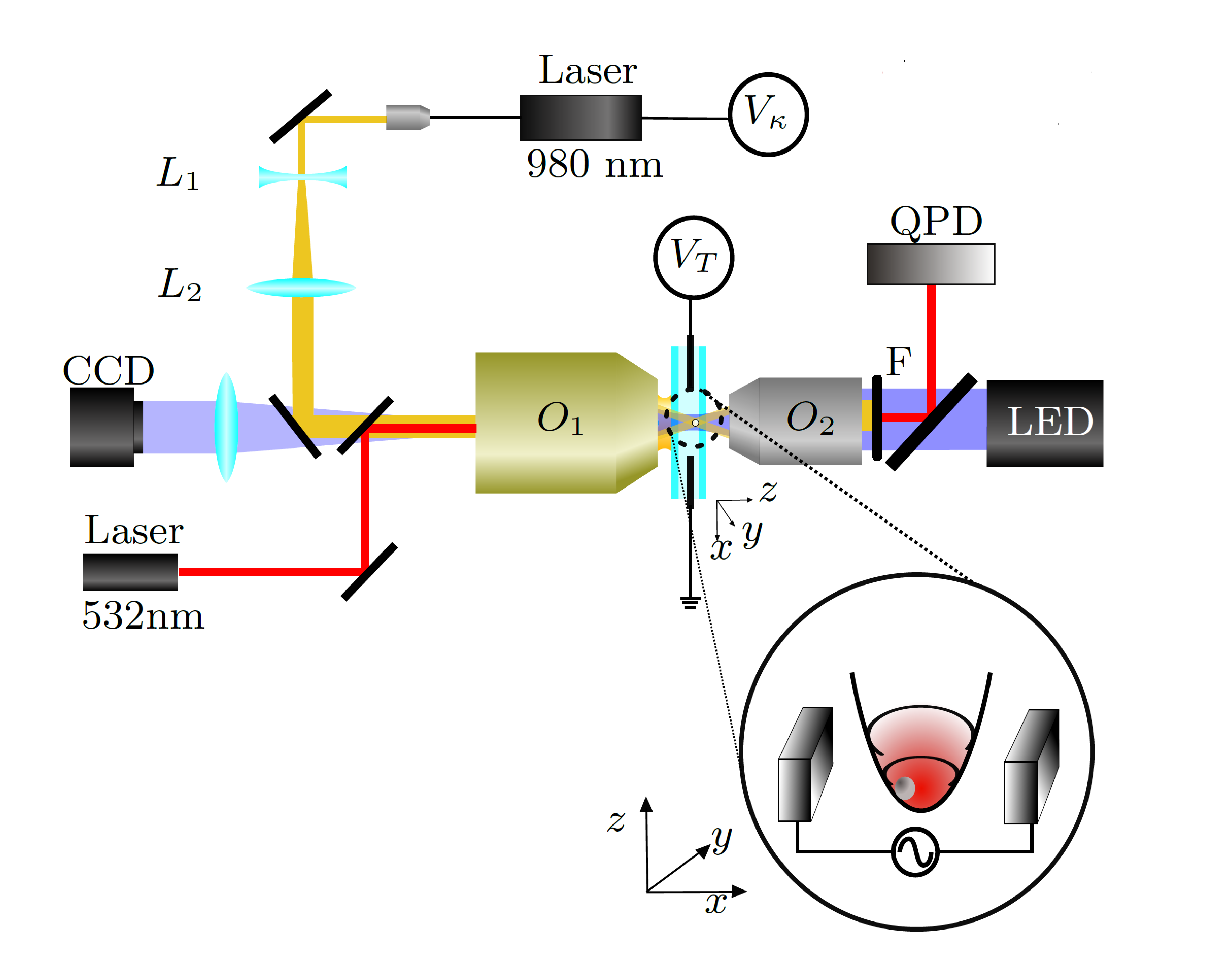} 
\caption{An optical trap is obtained by tightly focusing an infrared laser (980 nm) through the high NA objective $O1$. The stiffness of the trap is modified by a power supply that controls the intensity of the laser. A noisy electric potential $V_T$ is applied to the electrodes in the fluid chamber to control the kinetic temperature of the trapped particle. A red laser (532 nm) is used to track the position of the particle, by recording the forward scattered light collected by the objective $O2$ into a quadrant photodiode (QPD). A LED and a CCD camera are used for visualization. }
\label{fig:experiment}
\end{figure}

Figure~\ref{fig:experiment} shows a depiction of our experimental setup, which has been previously described~\cite{roldan2014measuring}. The setup is based on a horizontal self-built inverted microscope, where the sample is illuminated by a white lamp while the image is captured by a CCD camera. An infrared diode laser ($\lambda=980\,\rm nm$, Lumics, $100\,\rm mW$ maximum power) coupled in a single-mode fiber (Avanex, 1998PLM 3CN00472AG HighPower $980\,\rm nm$) is highly focused by a high numerical aperture (NA) immersion oil objective $O_1$ (Nikon, CFI PL FL 100$\times$ NA 1.30) to create the optical potential. Prior to entering the objective, the optical beam is expanded by lenses $L_1$ (focal length$=-30\,\rm cm$) and $L_2$ (focal length$=20\,\rm cm$) to overfill the input pupil of the objective. Laser controller (Arroyo Instruments 4210) allows the management of the optical power at a maximum rate of $250\,\rm kHz$ using an external voltage $V_{\kappa}$. Since the trap stiffness $\kappa$ depends linearly on the optical power, $\kappa$ can be controlled with at the same rate as the external voltage~\cite{mazolli2003theory}.

Polystyrene beads (G. Kisker-Products for Biotechnology, PPs-1.0, diameter $(1.00\pm0.05)\,\rm\mu m$) are diluted in Milli-Q water to a final concentration of a few microspheres per mL. The solution is injected into a custom-made fluid chamber, which is placed in a holder whose position in the three axes can be controlled with picomotors (Newport, 8752). Afterwards, the chamber is mechanically sealed to avoid fluxes and contamination, allowing us to work several days with the same solution. Polystyrene beads have an inherent charge in polar liquids which allows us to apply deterministic forces into our trapped microsphere. We add two aluminum electrodes at the two ends of the chamber to apply a controllable voltage ($V_T$) to the sample. Both $V_{\kappa}$ and $V_T$ are controlled by the same signal generator (Tabor electronics, WW5062) run by Labview software. In the case of $V_T$, the output signal of the signal generator is amplified $1000$ times with a high-voltage power amplifier (TREK, 623B).

The particle is tracked using an additional $532\,\rm nm$ laser collimated by a microscope objective ($\times$10, NA 0.10) and sent through the trapping objective ($O_1$). The light scattered by the trapped object is collected by the objective $O_2$ (Olympus, 40$\times$, NA 0.75) and projected into a quadrant photo detector (QPD, Newfocus 2911). The maximum acquisition frequency of the QPD is $200\,\rm kHz$. A $532\,\rm nm$ pass filter ($F$) blocks additional scattered light. The signal is transferred through an analog-to-digital conversion card (National Instruments PCI-6120) and recorded with LabView software.

The calibration of the nanodetection is obtained from the analysis of the thermal fluctuations of the bead within a static trap at room temperature. From the study of the power spectral density of the trajectories, both voltage-to-nanometers conversion factor, $S_{\rm QPD}$(nm/V), and  $\kappa$ are obtained. All experiments are done with the beads trapped 20 $\mu$m above the coverslip surface, in order to avoid surface effects in the friction coefficient $\gamma$~\cite{visscher1996construction}. The input voltage controls the noise intensity and can be linked to the effective temperature of the particle as $T_{\text{kin}}=T+S_T V^2_T$, where $S_{T}$(K/V$^2$) is the calibration factor. All calibrations are repeated each time a new bead is trapped. In the experiments presented here, $\kappa$ is calibrated as a function of $V_{\kappa}$ (data not shown),  $V_{\kappa}$ being of the order of $\rm pN/\mu m$, and noise amplitudes of the order of thousands of $V$ which led to values of $T_{\text{kin}}$ up to thousands of Kelvins. Note that, although we do not know the actual value of the electric field in our chambers, it is not needed for our calculations.

\section{Measurement of the  fluctuations of the instantaneous velocity}

The measurement of instantaneous velocity of a Brownian particle requires, in principle,  to sample the position of the particle with acquisition rates of the order of the momentum relaxation frequency $f_p=\gamma/2\pi m\sim\rm MHz$~\cite{kheifets2014observation} or use a different system where the friction is not as strong as it is in water~\cite{millen2013nanoscale}. We make use of another recent result described in Ref.~\cite{roldan2014measuring} that allows one to measure the mean squared instantaneous velocity at any time $t$ during a quasistatic process, $\langle v^2_t\rangle$ from the mean squared time averaged velocity, $\langle \overline{v}^2_t\rangle$, being the latter obtained from low frequency samplings ($f \sim \,\rm kHz$ or $\Delta t\sim \,\rm ms$), $\overline{v}_t =  \frac{1}{\Delta t} [x(t+\Delta t)-x(t)]$. For an underdamped Brownian particle of mass $m$ trapped with a quadratic potential of time-dependent stiffness $\kappa_t$, $U(x,t)=\frac{1}{2}\kappa_t x^2$ and immersed in a thermal bath at temperature $T$ , the mean squared instantaneous velocity equals to 
\begin{equation}
\langle v^2_t\rangle=\mathcal{L}_t \langle \overline{v}^2_t\rangle,
\end{equation}
 where $\mathcal{L}_t$ is a correction factor that depends on the acquisition frequency and on the physical parameters of the system at time $t$:
\begin{equation}
\mathcal{L}_t= \frac{1}{2f^2}\left[ \frac{1}{f_0^2} + \frac{e^{-\frac{f_p}{2f}}}{f_1} \left(  \frac{e^{-f_1/f}}{f_p+2f_1} - \frac{e^{f_1/f}}{f_p-2f_1}  \right)\right]^{-1},
\label{eq:FactorBdeC_trapped}
\end{equation}
where $f_0=\sqrt{f_p f_\kappa}$, $f_\kappa=\kappa_t/2\pi\gamma$ and $f_1=\sqrt{f_p^2/4-f_0^2}$. 

\begin{figure}
\includegraphics[width=6.5cm]{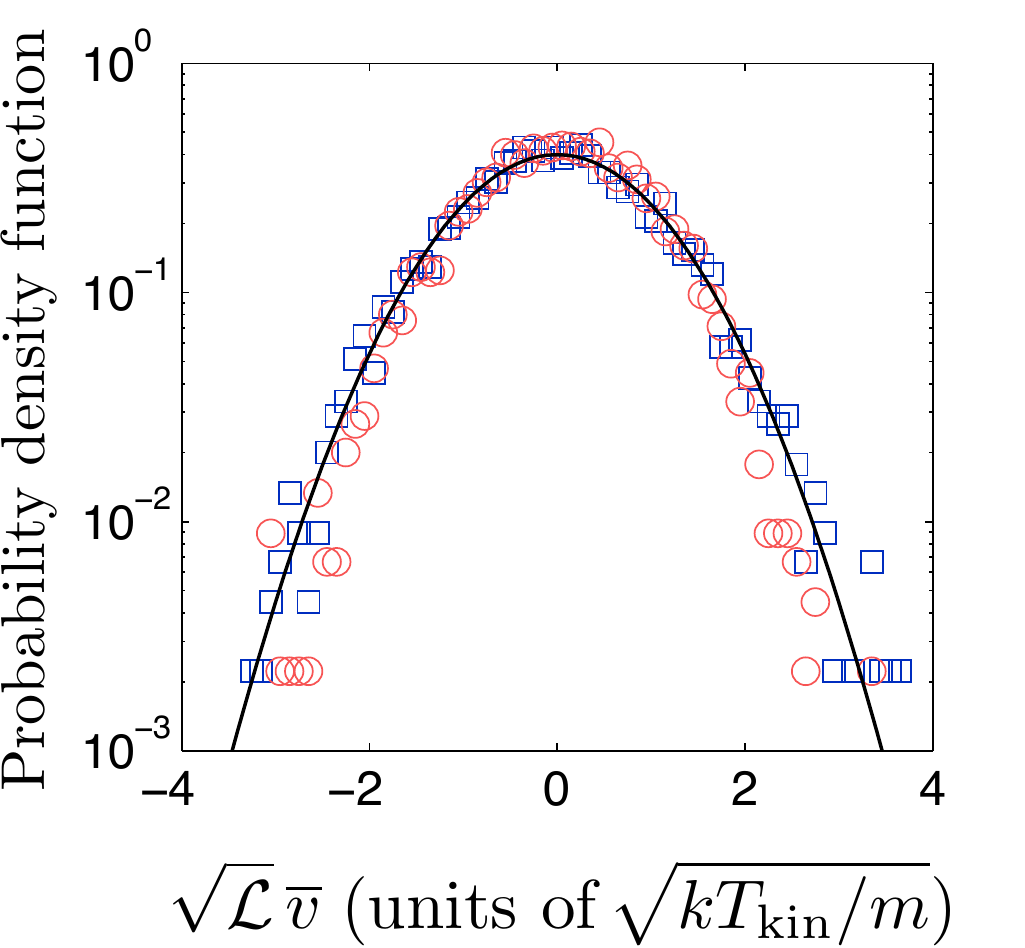} 
\caption{Probability density function of the time averaged velocity rescaled by $\sqrt{\mathcal{L}_t}$ at the beginning ($t=0$, blue squares) and at the end ($t=\tau=0.5\,\rm s$, red circles) of the pseudo-adiabatic process. The data of $\overline{v}_t$ is obtained from a time window $[t,t+S]$ with $S=5\,\rm ms$ for $t=0$ (beginning) and $t=0.5\,\rm s$ (end) for an ensemble of $900$ cycles of duration $\tau=0.5\,\rm s$. The rescaling is done using $\mathcal{L}_t$ calculated using Eq.~\eqref{eq:FactorBdeC_trapped}. We also show the Maxwell-Boltzmann equilibrium velocity distribution (black curve).}
\label{fig:PDFv}
\end{figure}

In a quasistatic process, $v_t$ is a Gaussian variable with zero mean and variance $\langle v_t^2\rangle=\frac{kT}{m}$. Since $\overline{v}_t = \frac{1}{\Delta t}\int_t^{t+\Delta t} v(s) ds$, the time-averaged velocity is also Gaussian distributed, as the sum of Gaussian variables is also a Gaussian variable. Its mean is trivially zero, $\langle \overline{v}_t\rangle=0$. To fully specify the distribution we must therefore only give its second momentum, which is $\langle v^2_t\rangle=\mathcal{L}_t \langle \overline{v}^2_t\rangle$ with the correcting factor $\mathcal{L}_t$ as given in \eqref{eq:FactorBdeC_trapped}. As a result, $\tilde v_t= \sqrt{L_t} \overline{v}_t$  has the same distribution as the instantaneous velocity $v$, or equivalenty,
\begin{equation}
\rho(\sqrt{L_t} \overline{v}_t) = \rho(v_t),
\label{eq:PDFveq}
\end{equation}
the latter being described by the equilibrium Maxwell-Boltzmann velocity distribution.

We tested the theoretical prediction given by Eq.~\eqref{eq:PDFveq} with data from the quasistatic processes described in the main text, as well as in the two control experiments (isothermal and isochoric processes). In Fig.~\ref{fig:PDFv} we show the probability density function of $\sqrt{\mathcal{L}_t} \overline{v}_t$ for $t=0$ and $t=\tau=0.5\,\rm s$ for the pseudo-adiabatic process, obtained from $900$ cycles. Both distributions coincide with Maxwell-Boltzmann distribution (black curve in Fig.~\ref{fig:PDFv}) with high accuracy for two orders of magnitude. The tails of the reconstructed experimental distributions deviate from the theoretical distribution due to statistical under sampling. This deviation might decrease when using data from a larger number of cycle repetitions.

\section{Averages of Work, Heat, Kinetic energy and Internal energy in quasistatic processes at the microscale}
\label{sec:energetics_theory}

Let us consider a Brownian particle of mass $m$ that is immersed in a fluid at temperature $T$. We assume that the particle moves in one dimension, and its position and velocity at time $t$ are denoted as $x_t$ and $v_t$, respectively. The particle is trapped with a potential $U(x_t,\lambda_t)$ that can be changed in time via a control parameter $\lambda_t$, and the temperature might change in time as $T_t$. The dynamics of such a Brownian particle is described by the underdamped Langevin equation~\cite{Langevin1908}
\begin{equation}
	m\frac{dv_t}{dt} = -\frac{\partial U(x_t,\lambda_t)}{\partial x}-\gamma v_t+ \xi_t + F_t,
	\label{eq:Langevin}
\end{equation}
where $\gamma$ is the friction coefficient of the particle in the fluid. Thermal fluctuations are modeled by a Gaussian white noise with $\langle \xi_t\rangle =0$ and $\langle \xi_t\xi_{t'}\rangle =2\gamma kT_t \delta(t-t')$, $k$ being Boltzmann's constant. The term $F_t$ accounts for any external forces that can be directly exerted to the particle. 

We consider quasistatic thermodynamic processes where the control parameter changes slower than any relaxation time of the system. In such a case, the phase space density of the system can be described by Gibbs distribution throughout the process. If the trapping potential is quadratic, $\kappa_t$ being the stiffness of the potential, $U(x_t,\lambda_t)=U(x_t,\kappa_t)=(1/2)\kappa_t x_t^2$, and the energy of the particle can be described by the Hamiltonian
\begin{equation}
\mathcal{H} (x_t,v_t,t) = \frac{1}{2}\kappa_t x^2_t + \frac{1}{2}mv^2_t.
\end{equation}
At time $t$, the system is described by a canonical state $\rho(x_t,v_t,t)=\exp[-\beta_t\mathcal{H}(x_t,v_t,t)]/Z_t$, where $\beta_t=1/kT(t)$ and $Z_t$ is the partition function, which is equal to
\begin{equation}
Z_t= \int \, dx\, dv\, e^{-\beta_t\mathcal{H}(x,v,t)}=  \left( \frac{4\pi^2 k^2}{m}\right)^{1/2}  \left(\frac{T^2_t}{\kappa_t} \right)^{1/2} .
\end{equation}
The free energy of the particle at time $t$ is
\begin{equation}
F_t=-kT_t \ln Z_t = -\frac{kT_t}{2} \ln \left( \frac{4\pi^2 k^2}{m} \frac{T^2_t}{\kappa_t}   \right).
\end{equation}
The entropy of the particle $S_t=-(\partial F_t / \partial T_t)$ satisfies $S_t \propto T^2_t/\kappa_t$. Therefore in a quasistatic adiabatic process $d(T^2_t/\kappa_t)=0$, or equivalently, $T^2_t/\kappa_t=\rm const$. The same analysis can be carried out using the overdamped description, i.e., neglecting the velocity degree of freedom. In such a case, it can be shown that $S_t \propto T_t/\kappa_t$, which implies that entropy in the position degree of freedom can be conserved when $T_t/\kappa_t=\rm const$.

\begin{table*}
\setlength{\tabcolsep}{12pt}
\centering
\ra{1.3}
\begin{tabular}{@{}llll@{}}\\ \hline 
Process &  $\langle W\rangle$ & $\langle Q_x\rangle$ & $\langle Q\rangle $ \\ \hline 
Isothermal &    $\frac{kT}{2} \ln \frac{\kappa_{\tau}}{\kappa_{0}}$ & $-\frac{kT}{2} \ln \frac{\kappa_{\tau}}{\kappa_{0}}$ & $-\frac{kT}{2} \ln \frac{\kappa_{\tau}}{\kappa_{0}} $ \\
Isochoric  &  $ 0 $ &  $\frac{k}{2}  (T_{\tau} - T_{0})$&$k (T_{\tau} - T_{0})$  \\
Pseudo-adiabatic &   $\frac{k}{2} (T_{\tau} - T_{0})$ & $0$ & $\frac{k}{2} (T_{\tau} - T_{0})$ \\
Adiabatic &   $k (T_{\tau} - T_{0})$ & $-\frac{k}{2} (T_{\tau} - T_{0})$& $0$  \\ \hline 
\end{tabular}
\caption{Protocol and theoretical values of the average work and heat done along the four different thermodynamic processes starting at $t=0$ and ending at $t=\tau$: Isothermal $T_t=\rm const$, isochoric $\kappa_t= \rm const$, pseudo-adiabatic $T_t/\kappa_t=\rm const$ and adiabatic $T_t^2/\kappa_t=\rm const$. $T_{0}$  ($T_{\tau}$) and $\kappa_{0}$ ($\kappa_{\tau}$) are the initial (final) values of the temperature and stiffness along the processes when any of the two parameters are changed in time. }
\label{tab:firstlaw}
\end{table*}

We now study thermodynamic processes where time runs in the interval $t\in [0,\tau]$. During such processes, the position and the velocity of the Brownian particle describe a trajectory $\{ v_t,x_t \}_{t=0}^{\tau}$. For systems described by an underdamped Langevin equation~\eqref{eq:Langevin}, the work exerted on the particle and the heat transferred from the thermal bath in an interval of time $[t,t+dt]$ are, respectively:
\begin{eqnarray}
 d'W_t   &=& \frac{ \partial U(x_t,\lambda_t)}{ \partial \lambda} \circ d\lambda_t \label{eq:WSeki},\\
 d'Q_t &=& (-\gamma v_t + \xi_t) \circ dx_t,   \label{eq:QxSeki}
\end{eqnarray}
where $\circ$ denotes the Stratonovich product~\cite{SekimotoBook2010}. Notice that using~\eqref{eq:Langevin}, the heat can be decomposed in two terms,
\begin{equation}
d'Q_t = \frac{m}{2}dv^2_t + \frac{ \partial U(x_t,\lambda_t)}{ \partial x_t} \circ dx_t,
\end{equation} 
the first term being the kinetic energy change $dE_{\rm kin,t} =  \frac{m}{2}dv^2_t$ and the second the heat transferred to the position of the particle, $d'Q_{x,t} = \frac{ \partial U(x_t,\lambda_t)}{ \partial x_t} \circ dx_t$. For a Brownian particle trapped with a harmonic potential of stiffness $\kappa_t$, equipartition theorem implies that $(1/2)\kappa_t\langle x_t^2\rangle = (1/2)kT_t $, and therefore, $\langle x_t^2\rangle = kT_t$ along the process, where the brackets denote ensemble average. In such a case, one can calculate the work transferred to the particle averaged over many realizations,
\begin{equation}
\langle W \rangle = \left\langle\int_0^t \frac{\partial U(x_t,\kappa_t)}{\kappa_t} d\kappa_t \right\rangle= \int\frac{1}{2}\langle x_t^2\rangle d\kappa_t = \int\frac{kT_t}{\kappa_t}d\kappa_t
\end{equation}
and the heat transferred to the position,
\begin{equation}
\langle Q_x \rangle = \left\langle\int_0^t \frac{\partial U(x_t,\kappa_t)}{x_t} \circ dx_t \right\rangle= \int\frac{\kappa_t}{2}d \langle x_t^2\rangle  = \int\frac{\kappa_t}{2}d\left(\frac{kT_t}{\kappa_t}\right).
\end{equation}
Notice that equipartition theorem implies that $\langle \Delta E_{\rm kin}\rangle =\langle \Delta U\rangle= \frac{k}{2}[T_{\tau}-T_{0}]$, $\Delta U$ being the internal energy change. The total energy change satisfies $\langle \Delta E\rangle = k[T_{\tau}-T_{0}]$. In Table~\ref{tab:firstlaw} we show the values of the quasistatic ensemble averages for selected thermodynamic processes: isothermal ($T_t=\rm const$), isochoric ($\kappa_t=\rm const$) and adiabatic processes. In the latter case, we distinguish between the pseudo-adiabatic process, where $T_t/\kappa_t=\rm const$, which yields $\langle Q_x\rangle=0$ but $\langle Q\rangle = \frac{k}{2} (T_{\tau} - T_{0})$, and the actual adiabatic process, where $\langle Q_x\rangle = -\frac{k}{2} (T_{\tau} - T_{0})$ and the total average heat vanishes $\langle Q\rangle = 0$.


\section{Distribution of the Energy change, Heat and Work in a non-isothermal quasistatic process}
\label{app:rhoq}

In this section, we calculate the energy change, heat and work distributions in quasistatic process in which a Brownian particle whose position is denoted as $x$ is trapped with a quadratic potential of stiffness $\kappa$, $U(x)=\frac{1}{2}\kappa x^2$. We consider both the overdamped description where only the position degree of freedom is taken into account, and the full underdamped description including the velocity.

 Along the protocol of duration $\tau$, the temperature changes from $T_0$ to $T_\tau$ and the stiffness from $\kappa_0$ to $\kappa_\tau$.

\subsection{Overdamped description}

 We assume that the process is quasistatic and therefore the distribution of the position at any time $t$ during the process is the equilibrium (Gaussian) distribution 
\begin{equation}
\rho (x,t)=\rho_{\rm eq}^{\{\beta_t, \kappa_t\}} (x) = \frac{e^{-\beta_t\kappa_tx^2 /2}}{Z_t^x},
 \label{eq:rhoeqx}
\end{equation}
where $\beta_t=1/kT_t$ and $Z_t^x=\sqrt{2\pi / \beta_t \kappa_t}$ is the partition function for the $x$ degree of freedom.

In a quasistatic process, the work distribution is peaked at its mean value~\cite{SekimotoBook2010}
\begin{equation}
\rho_W (W) = \delta (W-\langle W\rangle).
\end{equation}
Taking into account the First Law of Thermodynamics, $Q_x=\Delta U - W$, the heat distribution is equal to the distribution of the internal energy change centered in $\Delta U=-\langle W \rangle$
\begin{equation}
\rho_{Q_x}(Q_x)=\rho_{\Delta U}(\Delta U + \langle W\rangle).
\label{eq:rhoqdef}
\end{equation}
If the initial position of the particle is $x_0$ and the final position is $x_\tau$, the internal energy change is
\begin{equation}
\Delta U = \frac{1}{2}(\kappa_\tau x_\tau^2 - \kappa_0 x_0^2).
\end{equation}
We calculate the distribution of $\Delta U$ for a quasistatic process where the temperature and stiffness change from $(T_0,\kappa_0)$ to $(T_\tau,\kappa_\tau)$. The probability distribution of $\Delta U$ to be $\Delta U \in [u,u+du]$ is equal to $\rho_{\Delta U} (u) du$, where
\begin{eqnarray}
\rho_{\Delta U} (u) &=& \iint \delta \left( u- \frac{1}{2} (\kappa_\tau x_\tau^2 - \kappa_0 x_0^2) \right) \rho_{\rm eq}^{\{\beta_0,\kappa_0 \}} (x_0)\times\nonumber\\
&& \times\rho_{\rm eq}^{\{\beta_0,\kappa_0 \}}  (x_\tau)\, dx_0\,dx_\tau, \label{eq:rhoDUx}
\end{eqnarray}
where the integration is done from $-\infty$ to $\infty$ unless we specify different integration limits.

We now do the following change of variables
\begin{equation}
y_i=\frac{1}{2}\kappa_i x_i^2,
\end{equation}
for $i=0,\tau$. The equilibrium distribution of the random variable $y_i$ is
\begin{equation}
\rho_{\rm eq} ^{\{\beta_i,\kappa_i \}} (y_i) = \int \delta \left(y_i - \frac{1}{2}\kappa x_i^2 \right) \rho(x_i,\beta_i,\kappa_i) \,dx_i,
\label{eq:rhoydef}
\end{equation}
where the $\delta-$function in the integrand can be rewritten as
\begin{eqnarray}
 \delta \left(y_i - \frac{1}{2}\kappa x_i^2 \right) &=& \frac{1}{\sqrt{2\kappa_i y_i}} [ \delta(x_i-\sqrt{2y_i/\kappa_i}) \nonumber\\
 && + \delta(x_i+\sqrt{2y_i/\kappa_i}) ].
 \label{eq:deltasum}
\end{eqnarray}
By replacing~\eqref{eq:deltasum} and~\eqref{eq:rhoeqx} in~\eqref{eq:rhoydef}, and taking into account that $y_i$ can only take positive values,  we obtain
\begin{equation}
\rho_{\rm eq}^{\{\beta_i,\kappa_i \}} (y_i) = \sqrt{\frac{2}{\kappa_i y_i}} \frac{1}{Z_i} \, e^{-\beta_i y_i} \theta(y_i),
\label{eq:rhoy}
\end{equation}
where $\theta(y_i)$ is the step function evaluated at $y_i$, and $Z_i=\sqrt{2\pi / \beta_i \kappa_i}$.

The distribution of the internal energy change~\eqref{eq:rhoDUx} can be now expressed in terms of the new variables
\begin{eqnarray}
\rho_{\Delta U} (u) &=& \iint \delta \left( u-  y_\tau + y_0\right) \rho_{\rm eq}^{\{\beta_0,\kappa_0 \}} (y_0)\times\nonumber\\
 && \times\rho_{\rm eq}^{\{\beta_\tau,\kappa_\tau \}} (y_\tau)\, dy_0\,dy_\tau.
\end{eqnarray}
Integrating over $y_\tau$,
\begin{equation}
\rho_{\Delta U} (u) = \int  \rho_{\rm eq}^{\{\beta_0,\kappa_0 \}} (y_0) \rho_{\rm eq}^{\{\beta_\tau,\kappa_\tau \}} (u+y_0) \, dy_0.
\end{equation}
which yields, using the expression for the distribution $\rho_{\rm eq}^{\beta_i,\kappa_i}(y_i)$~\eqref{eq:rhoy},
\begin{equation}
\rho_{\Delta U} (u) = \frac{\sqrt{\beta_0 \beta_\tau}}{\pi}e^{-\beta_\tau u} \int \frac{e^{-(\beta_0 + \beta_\tau) y_0}}{\sqrt{y_0(u+y_0)}}\theta(y_0)\theta(u+y_0)\, dy_0. \label{eq:rhothetas}
\end{equation}
For $u>0$, the integral in~\eqref{eq:rhothetas} is equal to
\begin{equation}
\int_0^{\infty} \frac{e^{-(\beta_0 + \beta_\tau) y_0}}{\sqrt{y_0(u+y_0)}}\, dy_0 = e^{\frac{\beta_0+\beta_\tau}{2} u}\, K_0 \left(\frac{\beta_0+\beta_\tau}{2}u\right),
\end{equation}
where $K_0$ is the zeroth-order modified Bessel function of the second kind. For $u<0$,
\begin{equation}
\int_{-u}^{\infty} \frac{e^{-(\beta_0 + \beta_\tau) y_0}}{\sqrt{y_0(u+y_0)}}\, dy_0 = e^{\frac{\beta_0+\beta_\tau}{2} u}\, K_0 \left(-\frac{\beta_0+\beta_\tau}{2}u\right).
\end{equation}
Then, for any value of $u$, we obtain
\begin{equation}
\rho_{\Delta U} (u) =  \frac{\sqrt{\beta_0 \beta_\tau}}{\pi} e^{\frac{\beta_0-\beta_\tau}{2} u}\, K_0 \left(\frac{\beta_0+\beta_\tau}{2}|u|\right).
\label{eq:rhoU}
\end{equation}
The heat distribution~\eqref{eq:rhoqdef} is obtained from the distribution of the internal energy change~\eqref{eq:rhoU}, 
\begin{eqnarray}
\rho_{Q_x}(Q_x)&=&\frac{\sqrt{\beta_0 \beta_\tau}}{\pi} e^{\frac{\beta_0-\beta_\tau}{2} (Q_x+\langle W \rangle)}\times\nonumber\\ &&\times K_0 \left(\frac{\beta_0+\beta_\tau}{2}|Q_x+\langle W \rangle|\right) ,
\end{eqnarray}
which is Eq.~(1) in the Main Text. The distribution is asymmetric with respect to $Q_x=-\langle W\rangle$ except for the isothermal case ($\beta_0=\beta_\tau$).

\subsection{Underdamped case}
We start by computing the distribution of the energy change for a general non-isothermal quasistatic process. The total internal energy change is given by
\begin{equation}
\Delta E=\Delta U+\Delta E_\text{kin}=\frac{1}{2}(\kappa_\tau x_\tau^2 - \kappa_0 x_0^2)+\frac{1}{2}m(v_\tau^2 -v_0^2)
\end{equation}
Let us first derive the equilibrium distribution of the energy of state $i$, $E_i=\frac{1}{2}\kappa_i x_\tau^2+\frac{1}{2}mv_i^2$:
\begin{eqnarray}
\rho_{E_i}(E_i)&=&\frac{1}{Z_i^xZ_i^v}\int_{-\infty}^{\infty}\int_{-\infty}^{\infty}dx_idv_i e^{-\beta_\tau(\frac{1}{2}\kappa_i x_i^2+\frac{1}{2}mv_i^2)}\times\nonumber\\&& \times \delta(E_i-\frac{1}{2}\kappa_i x_i^2-\frac{1}{2}mv_i^2)
\end{eqnarray}
where the partition functions for the position and velocity degrees of freedom read $Z_i^x=\sqrt{\frac{2\pi}{\beta_i\kappa_i}}$ and $Z_i^v=\sqrt{\frac{2\pi}{\beta_i m}}$. We can first simplify the integral using the change of variables $u=\sqrt{\frac{\kappa_i}{2}}x_i, w=\sqrt{\frac{m}{2}}v_i$:
\begin{eqnarray}
\rho_{E_i}(E_i)&=&\frac{\beta_i}{\pi}\iint dudv e^{-\beta_i(u^2+w^2)}\times\nonumber\\&&\times \delta(E_i-u^2-w^2)
\end{eqnarray}
Transforming the integral to polar coordinates $R^2=u^2+w^2,\tan(\theta)=w/u$ and rewriting Dirac's delta as in \eqref{eq:deltasum}, we finally obtain an exponential distribution for the energy as expected for an equilibrium state:
\begin{equation}
\rho_{E_i}(E_i)=\beta_i e^{-\beta_i E_i}\theta(E_i).
\label{eq:dist_totalE}
\end{equation}
The step function $\theta$ reflects the fact that total energy is necessarily positive.

The distribution for the energy change can be computed as
\begin{eqnarray}
\rho_{\Delta E}(\Delta E)&=&\iint dE_0dE_\tau\beta_0\beta_\tau e^{-\beta_0E_0}e^{-\beta_\tau E_\tau}\times\nonumber\\ &&\times\theta(E_0)\theta(E_\tau) \delta(\Delta E-E_\tau+E_0)
\end{eqnarray}
which after using the delta to eliminate one integral can be readily shown to yield
\begin{equation}
\rho_{\Delta E}(\Delta E)=\left\{\begin{array}{c}
\frac{\beta_0\beta_\tau}{\beta_0+\beta_\tau}e^{-\beta_\tau\Delta E}, \text{ if }\Delta E\geq 0 \\
\frac{\beta_0\beta_\tau}{\beta_0+\beta_\tau}e^{\beta_0\Delta E}, \text{ if }\Delta E< 0 \\
\end{array}
 \right.
\label{eq:dist_deltaE}.
\end{equation}

Using the First Law and taking into account that work is delta distributed in a quasistatic process, we find that heat distribution is exponentially distributed, according to

\begin{equation}
\rho_{Q}(Q)=\left\{\begin{array}{c}
\frac{\beta_0\beta_\tau}{\beta_0+\beta_\tau}e^{-\beta_\tau (Q+\langle W\rangle)}, \text{ if }Q+\langle W\rangle\geq 0 \\
\frac{\beta_0\beta_\tau}{\beta_0+\beta_\tau}e^{\beta_0 (Q+\langle W\rangle)}, \text{ if }Q+\langle W\rangle < 0 \\
\end{array}
 \right.
\label{eq:dist_Qtotal}.
\end{equation}


Since the final result \eqref{eq:dist_deltaE} only depends on starting with an equilibrium distribution for the energy \eqref{eq:dist_totalE}, the total energy change distribution coincides with the work distribution for an adiabatic process in a classical Hamiltonian system as the one referred to in the text, where the heat is exactly zero for every trajectory:

\begin{equation}
\rho_{\mathcal{H}}(W)=\left\{\begin{array}{c}
\frac{\beta_0\beta_\tau}{\beta_0+\beta_\tau}e^{-\beta_\tau W}, \text{ if } W\geq 0 \\
\frac{\beta_0\beta_\tau}{\beta_0+\beta_\tau}e^{\beta_0 W}, \text{ if } W< 0 \\
\end{array}
 \right.
\label{eq:dist_W}.
\end{equation}

\section{Entropy fluctuations in a non-isothermal quasistatic process}

Total entropy production $\Delta S_{\rm tot}$ can be expressed as the sum of the system entropy change $\Delta S$ plus the entropy change in the environment, $\Delta S_{\rm env} = \int dQ/T$, where $Q$ is the total heat absorbed by the system. In the quasistatic limit, total entropy production vanishes and is delta distributed,
\begin{equation}
\rho_{\Delta S_{\rm tot}}(\Delta S_{\rm tot})=\delta(\Delta S_{\rm tot}).
\end{equation}
We might therefore consider fluctuations of system and environment entropies, which in this case satisfy $\Delta S_{\rm env} = -\Delta S$, and therefore,
\begin{equation}
\rho_{\Delta S_{\rm env}}(\Delta S) =\rho_{\Delta S}( -\Delta S).
\end{equation}

Let us consider the fluctuations of the system entropy change in a quasistatic thermodynamic process of duration $\tau$. The system entropy change from $t=0$ to $t=\tau$ is a state function and its value only depends on the initial and final micro state of the system, described by $\{x_0,v_0\}$ and  $\{x_\tau,v_\tau\}$, respectively,
\begin{equation}
\Delta S / k = \ln \frac{\rho(x_0,v_0,0)}{\rho(x_\tau,v_\tau,\tau)}.
\end{equation}
For a quasistatic process, initial and final distributions are canonical, yielding,
\begin{equation}
\Delta S / k = \ln \frac{e^{-\beta_0 \mathcal{H}(x_0,v_0,0)}/Z_0}{e^{-\beta_\tau \mathcal{H}(x_\tau,v_\tau,\tau)}/Z_\tau} = [\beta_\tau \mathcal{H}_\tau - \beta_0 \mathcal{H}_0] + \ln \frac{Z_\tau}{Z_0},
\label{eq:Sparts}
\end{equation}
which finally gives
\begin{eqnarray}
\Delta S / k &=& \frac{1}{2}\ln \left(\frac{T^2_\tau/\kappa_\tau}{T^2_0/\kappa_0} \right)+\\& +&  \frac{1}{2}\left[ \left(\beta_\tau \kappa_\tau x_\tau^2- \beta_0 \kappa_0 x_0^2 \right)  + m \left( \beta_\tau v_\tau^2- \beta_0 v_0^2       \right)       \right] \nonumber.
\label{eq:Ssys}
\end{eqnarray}
In the above formula, the first term is deterministic while the second is stochastic. The distribution of system entropy change is the distribution of the second term shifted by the value of the first term.

Computing the distribution of 
\begin{equation}
\Delta \Omega = \frac{1}{2}\left(\beta_\tau\kappa_\tau x_\tau^2 - \beta_0\kappa_0 x_0^2\right) +  \frac{m}{2}\left(\beta_\tau v_\tau^2 - \beta_0v_0^2\right) 
\end{equation}
follows exactly the same lines of the previous section computation for $\Delta E$, except $\beta$ factors have to be absorbed in the change of variables. Consequently, $\Delta \Omega$ also follows an exponential distribution:
\begin{equation}
\rho_{\Delta\Omega}(\Delta\Omega)=\frac{e^{-|\Delta \Omega|}}{2}
\end{equation}
Hence, system entropy is distributed according to
\begin{eqnarray}
\rho_{\Delta S/k}(\Delta S/k)&=&\rho_{\Delta\Omega}(\Delta S/k-\Delta S^\text{det}/k)=\nonumber\\&=&\frac{1}{2}e^{-|\Delta S/k-\Delta S^\text{det}/k|}\label{eq:Sfluct}
\end{eqnarray}
being $\Delta S^\text{det}/k=\frac{1}{2}\ln \left(\frac{T^2_\tau/\kappa_\tau}{T^2_0/\kappa_0} \right)$ the deterministic part of the entropy change.

For the overdamped case, 
\begin{equation}
\Delta S_x=\frac{1}{2}\left(\beta_\tau\kappa_\tau x_\tau^2 - \beta_0\kappa_0 x_0^2\right)+\frac{1}{2}\ln\left(\frac{T_\tau/\kappa_\tau}{T_0/\kappa_0}\right),
\end{equation}
and the derivation is analogous to that for the potential energy change $\Delta U$, also absorbing the $\beta$ factors in the change of variables. Thus, the entropy change for a process only considering the position degree of freedom follows a distribution given by
\begin{equation}
\label{eq:Sxfluct}
\rho_{\Delta S_x/k}(\Delta S_x/k)=\frac{1}{\pi}K_0(|\Delta S_x/k-\Delta S_x^{\rm det}/k|)
\end{equation}
with $\Delta S_x^{\rm det}/k=\frac{1}{2}\ln\left(\frac{T_\tau/\kappa_\tau}{T_0/\kappa_0}\right)$.

Since both distributions are even, the deterministic parts coincide with the mean value in both cases. When considering the distribution of system entropy shifted by its mean value $\Delta S-\langle \Delta S\rangle$, Eqs.~\eqref{eq:Sxfluct} and~\eqref{eq:Sfluct} yield Eqs.~(2) and~(3) of the Main Text, respectively.

\end{document}